\documentclass[aps,prl,twocolumn,groupedaddress,showpacs]{revtex4}

\usepackage{graphicx}

\begin{document}

\title{Spectrally narrow, long-term stable optical frequency reference based on a Eu$^{3+}$:Y$_{2}$SiO$_{5}$ crystal at cryogenic temperature}

\author{Qun-Feng Chen}
\email[]{Qun-Feng.Chen@uni-duesseldorf.de}
\author{Andrei Troshyn}
\author{Ingo Ernsting}
\author{Steffen Kayser}
\author{Sergey Vasilyev}
\author{Alexander Nevsky}
\author{Stephan Schiller}
\affiliation{Institut f\"{u}r Experimentalphysik, Heinrich-Heine-Universit\"{a}t D\"{u}sseldorf, 40225 D\"{u}sseldorf, Germany}

\date{\today}

\begin{abstract}
Using an ultrastable continuous-wave laser at 580 nm we performed spectral hole burning of Eu$^{3+}$:Y$_{2}$SiO$_{5}$ at very high spectral resolution. Essential parameters determining the usefulness as a ``macroscopic'' frequency reference: linewidth, temperature sensitivity, long-term stability were characterized, using a H-maser stabilized frequency comb. Spectral holes with linewidth as low as 6 kHz were observed and the upper limit of the drift of the hole frequency was determined to be on the order of 5$\pm$3 mHz/s. We discuss necessary requirements for achieving ultra-high-stability in laser frequency stabilization to these spectral holes.
\end{abstract}

\pacs{}

\maketitle

Frequency stabilized lasers are of importance for a variety of scientific and industrial applications. Present-day laser stabilization techniques utilize various types of frequency references e.g. atomic or molecular transitions, or modes of low-loss optical resonators. Another type of frequency reference is an ensemble of optical centers in a solid at cryogenic temperature. Appealing features of the latter are a relatively low sensitivity to environmental disturbances and simplicity of interrogation. 

Spectral hole burning (SHB) is well-established technique that allows to overcome limitations imposed by inhomogeneous broadening of absorption lines and to address narrow optical transitions of dopants in solids \cite{Macfarlane87,PhysRevB.70.214116}. The technique has been proposed and implemented for numerous applications, e.g. optical data storage and processing, quantum computing and laser stabilization \cite{Bottger:03,Pryde2002309,PhysRevB.63.155111}. Even if only one single transition of a particular dopant/host system is identified as suitable for frequency stabilization at a competitive level, this would suffice to provide frequency-stable radiation over essentially the complete optical range, since a femtosceond frequency comb, a virtual beat technique \cite{springerlink:10.1007/s003400100735}, and a nonlinear frequency conversion can be employed for transfer of the frequency stability to other frequencies.

In this work we study some fundamental properties of narrow persistent spectral holes in a particular system, Europium ions doped in an yttrium orthosilicate crystal. Our measurements are performed at very high resolution in the frequency domain, using, for the first time, to our knowledge, an ultra-stable and narrow-linewidth laser. We achieve long-lived holes with width as low as 6 kHz, the lowest linewidth reported so far for long-lived holes \cite{footnote1}.

The $^{7}F_{0}$ -- $^{5}D_{0}$ transition (580 nm) in Eu$^{3+}$:Y$_{2}$SiO$_{5}$ crystal exhibits one of the narrowest optical resonances in a solid and hence was thoroughly studied during past decades \cite{Yano:91,PhysRevLett.72.2179,PhysRevB.68.085109,Sellars2004150,Macfarlane2004310}. The $^{7}F_{0}$ and $^{5}D_{0}$ energy levels of Eu$^{3+}$ ion have no electronic magnetic moment. The atomic nuclei in Y$_{2}$SiO$_{5}$ crystal have small or no magnetic moment, therefore the contribution of the nuclear spin fluctuations to the homogeneous broadening of the transition is small. The homogeneous linewidth of 122 Hz ($2\times10^{-13}$ relative linewidth) was determined from photon echo decays. This is close to the 85 Hz limit imposed by the spontaneous emission. Persistent spectral holes with a lifetime of $\approx 100$ h occur because the excited $^{5}D_{0}$ state decays to long-lived hyperfine levels of the ground state. 

Studies of the Eu$^{3+}$:Y$_{2}$SiO$_{5}$ spectral holes using continuous-wave laser interrogation have been severely limited by the linewidth and frequency instability of the used laser sources, usually dye lasers. A solid-state laser source based on a cw-OPO has been used to obtain spectral holes with linewidth below 1 MHz \cite{Petelski:01}. By coincidence the 580 nm wavelength is very close to the $^{1}S_{0}$ -- $^{3}P_{0}$ clock transition in neutral Yb (578 nm). We used our diode-laser-based interrogation laser for an Yb optical lattice clock to significantly improve on previous measurements. 

\begin{figure}
\includegraphics[scale=0.5]{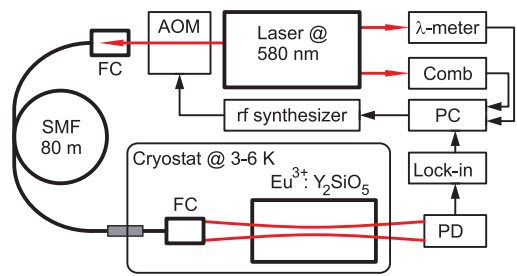}%
\caption{\label{fig:fig1}(color online). Experimental setup. AOM acoustooptic modulator, FC fiber collimator, SMF single mode optical fiber, PD photodetector.}
\end{figure}

A schematic of the experimental setup is shown in Fig.~\ref{fig:fig1}. Details of the narrowband laser are described in \cite{springerlink:10.1007/s00340-008-3113-4,Vogt2010}. In brief, we frequency double the IR output of the home-built external cavity diode laser in a quasi-phase-matched nonlinear waveguide. The laser is locked, via its second harmonic output, to a mode of a high-finesse Fabry-Perot resonator made of ultra low expansion glass (ULE), enclosed in a vacuum chamber. The setup is equipped with temperature stabilization and vibration isolation stages. The linewidth of the laser at 580 nm is about 1 Hz, the linear drift is $\approx0.1$ Hz/s. The laser frequency was measured by a Ti:Sapphire frequency comb referenced to a H-maser and to GPS. We used an acoustooptic frequency shifter (AOM) for tuning the laser frequency and to control the laser power. 

An uncoated $5\times5\times10$ mm$^{3}$ Y$_{2}$SiO$_{5}$ crystal doped with $0.1\%$ Eu$^{3+}$ (Scientific Materials) was cooled in a pulse tube cryostat to 3 -- 6 K. The laser and the cryostat, located in different laboratories, were connected by 80 m long single-mode optical fiber. Fluctuations in the unstabilized optical fiber resulted in broadening of the laser emission by $\approx1$ kHz. The maximum laser power delivered to the crystal was about 15 $\mu$W.

The laser beam was loosely focused in the crystal. The transmission through the crystal was detected by a low-noise silicon photodetector placed outside the cryostat. Spectral holes were burned on site 1 (516 847 GHz) during 2 -- 10 s at $\approx$~mW/cm$^{2}$ intensity and then the spectra were obtained by frequency-scanning the strongly (100 - 1000 times) attenuated laser relative to the hole spectrum, detecting the transmitted signal. The minimum delay between the burning and the reading was about 10 s.

\begin{figure}
\includegraphics[scale=0.35]{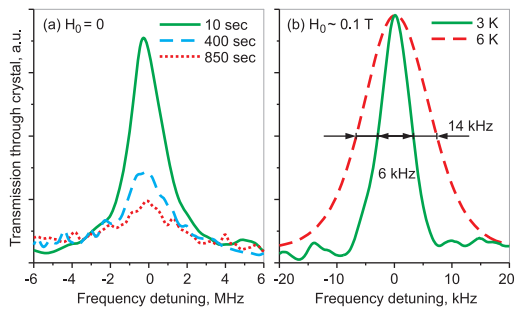}%
\caption{\label{fig:fig2}(color online). (a) Spectral holes obtained in the absence of the magnetic field at 6 K. Plot shows the spectra of the same hole shortly after burning (solid curve), 400 s after burning (dashed curve), and 850 s after burning (dotted curve). (b) Spectral holes obtained in $\approx0.1$ T magnetic field at 3 K (solid curve) and at 6 K (dashed curve).}
\end{figure}

Fig.~\ref{fig:fig2} illustrates spectral holes obtained in the absence of magnetic field ($H_{0} = 0$, left plot) and at $H_0 \approx 0.1$ T (here we placed the crystal inside a strong permanent magnet). At $H_{0} = 0$ and 6 K the width of spectral holes was 1.5 MHz a few seconds after the burning, and further broadened in time at a rate of 2 kHz/s. The holes became indistinguishable on the inhomogeneous background after a few tens of minutes. The introduction of the magnetic field resulted in two orders of magnitude sharper spectra (right plot). The width of the holes burned at 3 K was as low as 6 kHz, while burning at 6 K we found increased widths of 14 kHz. Furthermore, the spectral holes became much more stable in time as will be described below.

Our experiments reveal a considerable difference between the linewidths observed by other groups by photon echoes on a short time-scale ($\approx$~ms) and the widths observed in this work a few seconds after the hole burning. Instantaneous spectral diffusion (ISD) is not a likely reason of the hole's broadening in our experiments: the very small fraction of the exited ions ($\leq10^{-6}$ of the total Eu$^{3+}$ ions) results in a contribution $\Gamma_{ISD}\leq20$ Hz, according to \cite{PhysRevB.68.085109}.  The broadening of the spectral holes in time (continuous spectral diffusion) was reported for similar material (Eu$^{3+}$:Y$_{2}$O$_{3}$) \cite{Sellars:94}, and was attributed to an interaction with a distribution of low-lying energy states in the crystal that are thermally activated at low temperatures, similar to two-level systems (TLS's) in glasses. The same broadening mechanism can be expected in the Eu$^{3+}$:Y$_{2}$SiO$_{5}$ system. 


\begin{figure}
\includegraphics[width=7.5cm]{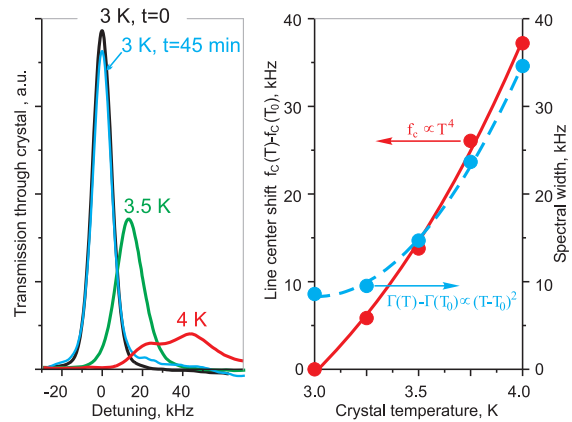}%
\caption{\label{fig:fig3}(color online). Left: spectra of the same hole (burned at 3 K) measured during the 3 K $\rightarrow$ 4 K $\rightarrow$ 3 K temperature cycling of the crystal. Right: cycles show measured temperature dependences of the hole's central frequency ($f_{c}(T)-f_{c}(T_{0})$, red) and linewidth ($\Gamma(T)-\Gamma(T_{0})$, blue), where $T_{0}$ is the temperature at which the hole burned. Solid lines show fit of the experimental data to power functions.}
\end{figure}

Another important characteristic of SHB frequency reference is its sensitivity to the temperature. We burned a spectral hole at 3 K and then measured the spectrum of this hole while the temperature of the crystal was increased incrementally to 4 K. At the end of the experiment the temperature was decreased back to 3 K and the hole's spectrum was measured again. The whole experiment took approximately 45 minutes. The slow drift of the laser frequency was monitored by the frequency comb during the measurements and was subtracted from the data. Fig.~\ref{fig:fig3}. left shows the spectrum of the original hole (black) and the spectra of the same hole at 3.5 and 4 K (green and red, respectively). The blue line shows the spectrum obtained after cooling the crystal back to 3 K. As can be seen, alteration of the crystal's temperature results in a shift of the hole's central frequency, broadening of the hole and deformation of its shape. However, all those effects are (almost) reversible, compare blue and black lines. The temperature cycling during 45 minutes resulted in a small deterioration of the hole's shape but no visible line broadening. Similar reversibility was observed when the hole was burned at 6 K and its spectra were read during the 6 K $\rightarrow$ 3 K $\rightarrow$ 6 K temperature cycling.

According to theory and measurement of K\"onz {\it et al.} \cite{PhysRevB.68.085109}, coupling to phonons causes the line center frequency and the linewidth of the SHB to change with  temperature as $f_{c}(T) \propto T^{4}$ and $\Gamma(T)-\Gamma(T=0) \propto T^{7}$, respectively. Our observations indicate that qualitatively $\Gamma(T) -\Gamma(T_{0}) \propto (T-T_0)^{2}$ (Fig.  \ref{fig:fig3} right, blue circles and dashed line), where $T_0$ is the temperature at which the hole is burned, and that the shape of the spectrum is deformed when the temperature difference is more than 1 K (Fig.  \ref{fig:fig3}, Left). The deformation implies an extra mechanism for broadening, e.g. interaction with TLS's or a (reversible) change in the magnetic field distribution within the crystal due to thermal expansion of the magnet and crystal housing. Our observation of the frequency shift agrees with a $T^{4}$ behavior (Fig. \ref{fig:fig3}, right, red circles and solid line). Because spectroscopy of a single hole is not suitable for covering a broad range of temperatures (the hole becomes too shallow and deformed for precise evaluation of $f_{c}$), we also measured the derivatives $\Delta f_{c}/\Delta T$ by repeating the process of burning a hole at certain temperature $T$ and comparing its spectra at the original and at the slightly shifted temperatures $T+0.25$ K. These measurements were carried out in the 3.75 - 5 K range and yielded $\Delta f_{c}/\Delta T \propto T^3$, in agreement with the previous result.

The measured temperature sensitivity $\Delta f_{c}/\Delta T \approx 20 $ Hz/mK at 3 K is very small.
With the crystal temperature stability we achieve now (2 $\mu$K) \cite{footnote2}
, the corresponding hole frequency instability should be reduced to below the $1\times10^{-16}$ level on a time scale of $10^{4}$ s.
A further strong reduction of the temperature sensitivity can be expected by cooling to the sub-K regime, due to the (predicted) $T^{3}$ dependence of $df_{c}/dT$.
 
\begin{figure}
\includegraphics[scale=0.35]{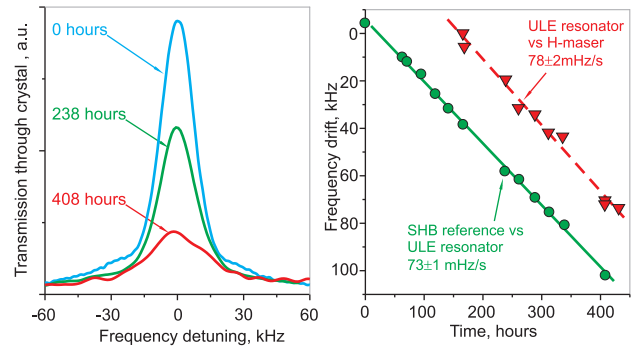}%
\caption{\label{fig:fig4}(color online). Left: spectra of the same hole obtained right after burning, 238 h, and 408 h after burning. The center frequencies have been overlaid. Right: drift of the frequency difference between ULE resonator frequency and H-maser frequency (red triangles) and drift of frequency difference between SHB reference and ULE resonator frequency (green circles).}
\end{figure}

To characterize the lifetime and the long-term stability of the spectral holes in Eu$^{3+}$:Y$_{2}$SiO$_{5}$ crystal we burned a hole at 3 K and then measured the spectra of this hole daily for two weeks. The temperature of the crystal was stable within 1 mK during this experiment. The shape of the hole remained very constant in time as illustrated in Fig.~\ref{fig:fig4} (left plot). The hole's average broadening rate of 0.85 kHz/day and decay rate of $4\%$ /day were estimated from a linear fit of the experimental data. The broadening varied strongly, accelerating up to few kHz/day and then slowing down and even reversing during the next days. This peculiar behavior is probably due to fluctuations of the stray fields inside the cryostat or microscopic displacements of the crystal (the crystal was loosely mounted inside the magnet to avoid strain; furthermore our pulse tube cooler produces vibrations with $\approx1$ $\mu$m amplitude at 1 Hz).

The long-term frequency stability of the SHB reference was evaluated by measuring the frequency difference with a cavity mode, $f_{c}-f_{\rm ULE}$, and the frequency difference $f_{\rm ULE}-f_{\rm maser}$. These differences were fitted to a linear dependence, and the difference of the slopes yields the average drift of the hole in absolute terms, 5$\pm$3 mHz/s. This is significantly smaller than the drift of the ULE cavity (76 mHz/s).

Laser frequency stabilization to a SHB reference has been demonstrated for a number of material systems \cite{Bottger:03,Pryde2002309,PhysRevB.63.155111}, where phase-modulation methods were used to lock the laser to the SHB reference. A SHB reference differs from other references in that its spectrum is modified in time by the probe laser radiation (and due to the spectral diffusion). Roll-off of the reference due to the interaction with the laser imposes a limit on long-term stability of a frequency-locked laser \cite{Julsgaard:07,Pryde2001587}.

The combination of a SHB reference with a laser prestablized to a high-finesse reference resonator could provide an efficient way to overcome this difficulty. A long-lived SHB reference can then be used for compensation of the slow drift of the resonator (or of other standard references, e.g. molecular iodine gas).

A first test of frequency stabilization of our prestabilized laser (where we achieved an instability below the 10$^{-13}$ level) has indicated several basic but important requirements in order to reach competitive results: stabilization of the optical path between laser and crystal, stabilization of the temperature of the crystal to the $\mu$K level (already achieved), and sufficient laser power for obtaining a sufficiently high signal-to-noise ratio. In addition, interrogation parameters (duty cycle, interrogation intensity, laser beam size,  crystal thickness, dopant concentration, etc.) should be optimized with respect to the magnitude of the drift to be corrected, in order to achieve optimum performance.  

Beyond these parameters, the inhomogeneous broadening ($\approx2$ GHz for the employed crystal) offers additional potential for improving the signal-to-noise ratio of the SHB reference \cite{Bottger:03}. Within this linewidth $\approx10^{4}$ -- $10^{5}$ frequency channels are in principle accessible. The number of spectral holes (channels), which can be burned simultaneously, is limited essentially by the available laser power. In our experiments we produced single spectral holes by burning at a few $\mu$W power during several seconds. Thus, with several 10 mW, a realistically attainable power, a number of holes of the above order can indeed be simultaneously burnt. This could be implemented by producing a laser spectrum that is amplitude-modulated at a large number of discrete frequencies by means of a low-voltage wide-band fiber-optic modulator. The simultaneous read-out could be performed with an appropriate modulation/demodulation scheme. The use of multiple SHB references for laser stabilization allows reducing the exposure of each individual reference to the laser radiation and hence reducing its spectrum's distortion, while delivering a sufficient total laser power to the transmission detector.

With our current setup, the signal-to-noise ratio of the spectrum is about $10^{3}$ for an integration time of 0.1 s. Therefore, if 100 holes were used simultaneously, a frequency instability of about 0.1 Hz ($2\times10^{-16}$) at an integration time of 10 s could be expected. This frequency instability would be competitive with the currently best room temperature ULE reference cavity \cite{Jiang:2011:158}.

In summary, we demonstrated narrow, long-lived spectral holes in Eu$^{3+}$:Y$_{2}$SiO$_{5}$ crystals using, for the first time, to our knowledge, an ultra-stable and narrow-linewidth laser. The narrow linewidth (6 kHz), low temperature sensitivity (20 Hz/mK) very good long-term stability (5$\pm$3 mHz/s drift) of the Eu$^{3+}$:Y$_{2}$SiO$_{5}$ SHB reference at 3 K in moderate magnetic field ($\approx0.1$ T) represent a very attractive combination of properties for laser frequency stabilization, especially in conjunction with prestabilization to an optical cavity. One can expect a further improvement of those parameters at temperatures below 3 K and in stronger magnetic fields. High-resolution SHB in the frequency domain may also be a useful tool for studying the nature of spectral diffusion mechanisms in solids.

The authors are very grateful to Sirah Laser- und Plasmatechnik GmbH  for loan of the single-frequency dye laser.

{\it After completion of this manuscript, we learned of the work by Thorpe et al. (arXiv:1106.0520v1) on frequency stabilization of a dye laser to a level of $6\times10^{-16}$ on time scales between 2 and 8 s via SHB of Eu$^{3+}$:Y$_2$SiO$_5$}.


\begin{thebibliography}{20}
\bibitem[{foo({\natexlab{a}})}]{footnote1}
{\bibinfo{title}{Spectral holes with 3.5 kHz width were reported for a
  similar material, Eu$^{3+}$:Y$_{2}$O$_{3}$ \cite{Sellars:94}. however, these
  spectra were measured in frequency domain shortly after burning (10 ms) and
  the burning itself occurred during 2 ms}}.
\bibitem[{foo({\natexlab{b}})}]{footnote2}
{\bibinfo{title}{A temperature stability of 2 $\mu$K was achieved in a
  time scale of $10^{4}$ s by active stabilization, using carbon-glass resistor
  for temperature measurement and a commercial resistance bridge temperature
  controller, with feedback to a heater close to the crystal}}.
\end{thebibliography}

\end{document}